\documentclass[12pt]{article}
\usepackage{lingmacros}
\usepackage[english]{babel}
\usepackage{graphicx}

\large \baselineskip = 20 true pt

\begin{document}
\begin{center}
{\Large \bf Use the Keys Pre-Distribution KDP-scheme for Mandatory Access Control Implementation } \vspace{0.5cm}
\end{center}

\begin{center}
S.V. Belim, S.Yu. Belim \\
Dostoevsky Omsk State University, Omsk, Russia
 \vspace{0.5cm}
\end{center}

\begin{center}
{\bf Abstract}
\end{center}

{\small
The possibility of use the keys preliminary distribution KDP-scheme for mandatory access control
realization in the distributed systems with user's hierarchy is considered. The modified keys
preliminary distribution algorithm is suggested. It is developed a method for creation of subsets
family for solution this task.
}


\section{Introduction}

Mandatory differentiation of access is more rigorous in comparison with a discretionary analog. 
The centralized security subsystem is necessary for its realization. At system there has 
to be a uniform center of a decision making comparing mandates of access. This problem is easily
solved in local systems. There are some difficulties for the distributed systems. 
Now this problem is solved on the basis of open keys certificates. Such decision cannot 
be considered satisfactory. Certificates use sluggish asymmetric cryptographic algorithms. 
The system based on certificates uses the center of confirmation.

Qualitatively other algorithm can be constructed on the basis of keys preliminary distribution
schemes. In this case the role of the central server comes down only to key materials 
distribution. Network's subscribers calculate keys of information exchange self-contained. 
The main problem consists that widely known keys preliminary distribution schemes \cite{b1,b2}
provide information exchange for each user with everyone. Modification of such schemes 
is necessary for accounting of security policy of system. Modifications of the Blom's keys
preliminary distribution scheme, considering the forbidden channels it is suggested in work 
\cite{b3}. The organization of simplex channels for the same scheme is realized in article
\cite{b4}. The solution similar task on the basis of the KDP scheme is proposed in articles 
\cite{b5,b6}.  These works are focused on realization of discretionary security policy. 
Mandatory security policy demands accounting of hierarchy, both subjects, and objects. 
The decision on the basis of hash functions is suggested in article \cite{b7}. However 
this approach does not allow realizing exchange between users taking into account hierarchy.

The purpose of this article is development of the keys preliminary distribution scheme allowing
realizing mandatory security policy in the distributed computing systems.

\section{Keys preliminary distribution scheme}

Mandatory access control uses a security tags set which form an algebraic lattice. Security 
tags are both at users, and at informational objects. At request for access there is a comparison
of security tags. The decision is made on the basis of some logical condition.

Let's designate the set of users in the distributed system $U$. For users of system there is 
an order relation. We will be limited to the order relation described by the graph in the form 
of a tree. 	Dominance of the user $u_i$ over the user $u_j$ we will designate $u_i>u_j$. 
Also the situation when users are incomparable with each other is possible. Let's use then
designation $u_i<>u_j$. We will be limited to a case of mandatory access control in which only
informational streams from below up are resolved. This case corresponds to mandatory security
policy on ensuring confidentiality of information. In this case for two users $u_i>u_j$  
is resolved only informational stream from $u_j$ to $u_i$. For two incomparable users 
informational streams in both parties are forbidden.

Let's set the task to formation the key scheme allowing communicating according to mandatory 
access control. For this purpose we will construct the keys preliminary distribution scheme
calculating pair keys only for the allowed channels.

For the solution this task we modify the KDP scheme of keys preliminary distribution. 
For system without access control in the KDP scheme key materials is formed in based of set
$K=\{k_1,...,k_n\}$. Key materials beforehand are sent to all users via secure channels.  
For development of pair keys the system subsets $S=\{S_1,...,S_m\}$ of set $\{1,...,n\}$ 
is used. $m$ -- number of users in system. The set $S$ is open. For information exchange with 
the user $u_j$ the user  $u_i$ takes subsets $S_i$ и $S_j$. Further he calculates the elements
entering in the product of sets $S_{ij}=S_i\cap S_j$. The pair key is calculated with use the
key materials $K$, and subsets $S_{ij}$:
\[
k_{ij}= \oplus k_l\ \ \ (l\in S_{ij}).
\]
The same operations are carried out by the user $u_j$ when obtaining the message from $u_i$.

The scheme described above allows carrying out exchange of messages for each user with everyone 
in both directions. We modify the scheme, having entered into it asymmetry of keys 
$k_{ij}\neq k_{ji}$. For this purpose also we use the key materials $K$ and a set $S$.  
For calculation the key of encrypting for the channel from $u_j$ to $u_i$ we use the difference 
of two sets:
\begin{eqnarray}
&& \triangle S_{ij}=\left\{ \begin{array}{c}
                              S_i\setminus S_j,\ if \ S_i\cap S_j=\emptyset  \\
                              S_i\cap S_j,\ overwise
                            \end{array}
\right.\nonumber\\
&& k_{ij}=\bigoplus_{l\in S_{ij}} k_l.\nonumber
\end{eqnarray}

Such approach leads to automatic implementation the requirement of  keys asymmetry. 
For reading messages the user $u_i$  ($i=1,...,m$) will use keys $k_{ij}$ ($j=1,...,m$), 
and for sending messages -- keys $k_{ji}$ ($j=1,...,m$). The suggested scheme is based on 
the symmetric encrypting that accelerates processes of encrypting and decrypting.

We realize the ban on channels of information exchange. For this purpose we will demand that 
the corresponding pair keys were zero $k_{ji}=0$, that is  $S_{ij}=\emptyset$. From here we 
receive requirements to a set of subsets $S$. The most widespread approach to creation the set 
of $S$ is uses of the Sperner's families \cite{b1}. The Sperner's family \cite{b2} is called 
the family of subsets $D=\{D_1,...,D_n\}$ such that, if $D_i\cap D_j\subseteq D_t$, that either
$t=i$, or $t=j$. In the unmodified KDP scheme on the basis of elements $D_i$ the Shperner's 
family are formed $S_{ij}$. We use similar approach for the solution the problem. Let's create 
the Shperner's family with the quantity of elements equal to number of users 
$D=\{D_1,...,D_m\}$. We will form a set $S$, moving on a tree of users hierarchy leaves 
to a root. Let's allocate "leave's" users $u_1,..., u_l$, where $l$ -- quantity of leave's tops 
on the tree. Let's equate, the elements of a set $S$ corresponding to them, to Sperner's family
elements $S_i=D_i$ ($i=1,...,l$). Let's rise from leaves to a tree root. If the top of $u_i$ has
the closest descendants $u_{i1},..., u_{ik}$, then to this user there corresponds the set:
\[
S_i=S_{i1}\cup S_{i2}\cup ... S_{ik}\cup D_i.
\]
This algorithm of formation the set $S$ leads to realization of the required condition the
mandatory access control: if $u_i>u_j$, then $S_i\supset S_j$, and 
$S_i\setminus S_j\neq \emptyset$, but $S_j\setminus S_i=\emptyset$. Thus, users can create a pair
key only for the allowed communication channels. Also the requirement for incomparable users 
is fulfilled: if $u_i<>u_j$, then $S_j\setminus S_i=\emptyset$  and $S_i\setminus S_j=\emptyset$.

\section{Example of keys preliminary distribution scheme}

Let's consider implementation the suggested scheme on a simple example. In system seven users are authorized. The hierarchy of users is shown in the figure 1.

\begin{figure}[ht]
\centering
\includegraphics[width=0.5\textwidth]{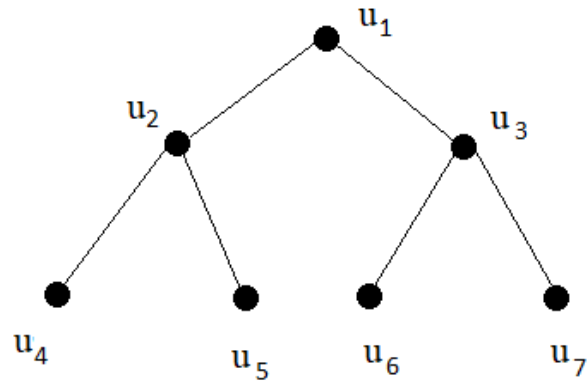}
\caption{Hierarchy of users.}
\label{fig1}
\end{figure}

Let's define the key materials set with 15 one byte elements:
\begin{eqnarray}
&&k_1= 00100100,\	\ k_2= 10101010,\ \	k_3= 01010101,\ \	k_4= 11011011,	\nonumber\\
&&k_5= 11101110,\ \ k_6= 00010001,\ \	k_7= 10010010,	k_8= 10110110,	\nonumber\\
&&k_9= 00011000,\ \ 	k_{10}= 11101110,\ \ k_{11}= 10111001,\ \ k_{12}= 11100111,\nonumber\\	
&&k_{13}= 00101101,\ \	k_{14}= 11010010,\ \	k_{15}= 01111111.\nonumber
\end{eqnarray}
We will set the Sperner's family as follows:
\[
D_1=\{1,2\},\ D_2=\{3,4\},\ D_3=\{5\},\ D_4=\{6,7,8\},
\]
\[
D_5={9,10},\ D_6=\{11,12,13\}, \ D_7=\{14,15\}.
\]
For sheet tops the sets S are defined as
\[
S_4=D_4=\{6,7,8\},\ \ S_5=D_5=\{9,10\},\ \ S_6=D_6=\{11,12,13\},\ \ S_7=D_7=\{14,15\}.
\]
For other users
\begin{eqnarray}
&&S_2=S_4\cup S_5 \cup D_2=\{3,4,6,7,8,9,10\},\ \
S_3=S_6\cup S_7\cup D_3=\{5,11,12,13,14,15\},\nonumber\\
&&S_1=S_2\cup S_3\cup D_1=\{1,2,3,4,5,6,7,8,9,10,11,12,13,14,15\}.\nonumber
\end{eqnarray}
Process of formation sets $S$ is presented in the figure 2.

\begin{figure}[ht]
\centering
\includegraphics[width=0.5\textwidth]{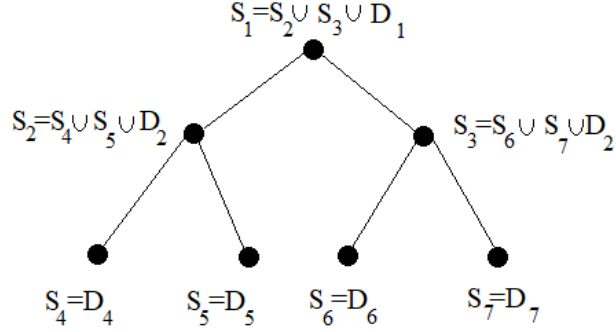}
\caption{Formation sets $S$.}
\label{fig2}
\end{figure}

For calculation of pair keys we will define set differences
\begin{eqnarray}
&& \Delta S_{12}=S_1\setminus S_2=\{1,2,5,11,12,13,14,15\},\nonumber\\
&& \Delta S_{13} = S_1\setminus S_3=\{1,2,3,4,6,7,8,9,10\},\nonumber\\
&& \Delta S_{14} = S_1\setminus S_4=\{1,2,3,4,9,10,11,12,13,14,15\},\nonumber\\
&& \Delta S_{15} = S_1\setminus S_5=\{1,2,3,4,5,6,7,8,11,12,13,14,15\},\nonumber\\
&& \Delta S_{16} = S_1\setminus S_6=\{1,2,3,4,5,6,7,8,9,10,14,15\},\nonumber\\
&& \Delta S_{17} = S_1\setminus S_7=\{1,2,3,4,5,6,7,8,9,10,11,12,13\},\nonumber\\
&& \Delta S_{21} = S_2\setminus S_1=\emptyset, 
\Delta S_{23} = S_2\setminus S_3=\emptyset,\nonumber\\
&& \Delta S_{24} = S_2\setminus S_4=\{3,4,9,10\},\nonumber\\
&& \Delta S_{25} = S_2\setminus S_5=\{3,4,6,7,8\},\nonumber\\
&& \Delta S_{26} = S_2\setminus S_6=\emptyset, 	 
\Delta S_{27} = S_2\setminus S_7=\emptyset, 	\nonumber\\
&& \Delta S_{31} = S_3\setminus S_1=\emptyset, \Delta S_{32} = S_3\setminus S_2=\emptyset, 	 
\Delta S_{34} = S_3\setminus S_4=\emptyset, 	
\Delta S_{35} = S_3\setminus S_5=\emptyset,\nonumber\\
&& \Delta S_{36} = S_3\setminus S_6=\{5,14,15\},\nonumber\\
&& \Delta S_{37} = S_3\setminus S_7=\{5,11,12,13\},\nonumber\\
&& \Delta S_{41} = S_4\setminus S_1=\emptyset, \Delta S_{42} = S4\setminus S2=\emptyset, 	 
\Delta S_{43} = S_4\setminus S_3=\emptyset, 	
\Delta S_{45} = S_4\setminus S_5=\emptyset,\nonumber\\	
&& \Delta S_{46} = S_4\setminus S_6=\emptyset,\ \	 
\Delta S_{47} = S_4\setminus S_7=\emptyset,\nonumber\\
&& \Delta S_{51} = S_5\setminus S_1=\emptyset,\ \ 
\Delta S_{52} = S_5\setminus S_2=\emptyset,\ \ 	 
\Delta S_{53} = S_5\setminus S_3=\emptyset,\ \ 	
\Delta S_{54} = S_5\setminus S_4=\emptyset,\nonumber\\
&& \Delta S_{56} = S_5\setminus S_6=\emptyset,\ \ 
\Delta S_{57} = S_5\setminus S_7=\emptyset,\nonumber\\
&& \Delta S_{61} = S_6\setminus S_1=\emptyset,\ \  	 
\Delta S_{62} = S_6\setminus S_2=\emptyset,\ \
\Delta S_{63} = S_6\setminus S_3=\emptyset,\ \ 	
\Delta S_{64} = S_6\setminus S_4=\emptyset,\nonumber\\
&& \Delta S_{65} = S_6\setminus S_5=\emptyset,\ \ 
\Delta S_{67} = S_6\setminus S_7=\emptyset,\nonumber\\
&& \Delta S_{71} = S_7\setminus S_1=\emptyset,\ \ 
\Delta S_{72} = S_7\setminus S_2=\emptyset,\ \ 	 
\Delta S_{73} = S_7\setminus S_3=\emptyset,\ \ 	
\Delta S_{74} = S_7\setminus S_4=\emptyset,\nonumber\\
&& \Delta S_{75} = S_7\setminus S_5=\emptyset,\ \
\Delta S_{76} = S7\setminus S6=\emptyset.\nonumber
\end{eqnarray}
Pair keys will be defined by equalities:
\begin{eqnarray}
&&k_{12}= k_1\oplus k_2\oplus k_5\oplus k_{11}\oplus k_{12}\oplus k_{13}\oplus k_{14}\oplus k_{15}=10111110,\nonumber\\
&&k_{13}= k_1\oplus k_2\oplus k_3\oplus k_4\oplus k_6\oplus k_7\oplus k_8\oplus k_9\oplus k_{10}=11000011,\nonumber\\
&&k_{14}= k_1\oplus k_2\oplus k_3\oplus k_4\oplus k_9\oplus k_{10}\oplus k_{11}\oplus k_{12}\oplus k_{13}\oplus k_{14}\oplus k_{15}=00101000,\nonumber\\
&&k_{15}= k_1\oplus k_2\oplus k_3\oplus k_4\oplus k_5\oplus k_6\oplus k_7\oplus k_8\oplus k_{11}\oplus k_{12}\oplus k_{13}\oplus k_{14}\oplus k_{15}=00000101,\nonumber\\
&&k_{16}= k_1\oplus k_2\oplus k_3\oplus k_4\oplus k_5\oplus k_6\oplus k_7\oplus k_8\oplus k_9\oplus k_{10}\oplus k_{14}\oplus k_{15}=10000000,\nonumber\\
&&k_{17}= k_1\oplus k_2\oplus k_3\oplus k_4\oplus k_5\oplus k_6\oplus k_7\oplus k_8\oplus k_9\oplus k_{10}\oplus k_{11}\oplus k_{12}\oplus k_{13}=01011110,\nonumber\\
&&k_{21}=0,\ \ 	k_{23}=0,\nonumber\\
&&k_{24}= k_3\oplus k_4\oplus k_9\oplus k_{10}=01111000,\nonumber\\
&&k_{25}= k_3\oplus k_4\oplus k_6\oplus k_7\oplus k_8=10111011,\nonumber\\
&&k_{26}=0,\ \ 	k_{27}=0,\nonumber\\
&&k_{31}=0,\ \ 	k_{32}=0,\ \ 	k_{34}=0,\ \ 	k_{35}=0,\nonumber\\
&&k_{36}= k_5\oplus k_{14}\oplus k_{15}=01100011,\nonumber\\
&&k_{37}= k_5\oplus k_{11}\oplus k_{12}\oplus k_{13}=10011111,\nonumber\\
&&k_{41}=0,\ \ 	k_{42}=0,\ \ 	k_{43}=0,\ \ 	k_{45}=0,\ \ 	k_{46}=0,\ \ 	k_{47}=0,\nonumber\\
&&k_{51}=0,\ \	k_{52}=0, \ \	k_{53}=0,\ \ 	k_{54}=0,\ \ 	k_{56}=0,\ \ 	k_{57}=0,\nonumber\\
&&k_{61}=0,\ \ 	k_{62}=0, \ \	k_{63}=0,\ \ 	k_{64}=0,\ \ 	k_{65}=0,\ \ 	k_{67}=0,\nonumber\\
&&k_{71}=0,\ \ 	k_{72}=0,\ \ 	k_{73}=0,\ \ 	k_{74}=0,\ \ 	k_{75}=0,\ \ 	k_{76}=0.\nonumber
\end{eqnarray}
The constructed keys preliminary distribution scheme satisfies the hierarchy of subjects shown in
the figure 1. Only information channels are resolved "from below-up". Incomparable users also
cannot communicate.

The suggested scheme can be used also in systems with the hierarchy of users other than a tree. Let's review an example in the figure 3.

\begin{figure}[ht]
\centering
\includegraphics[width=0.5\textwidth]{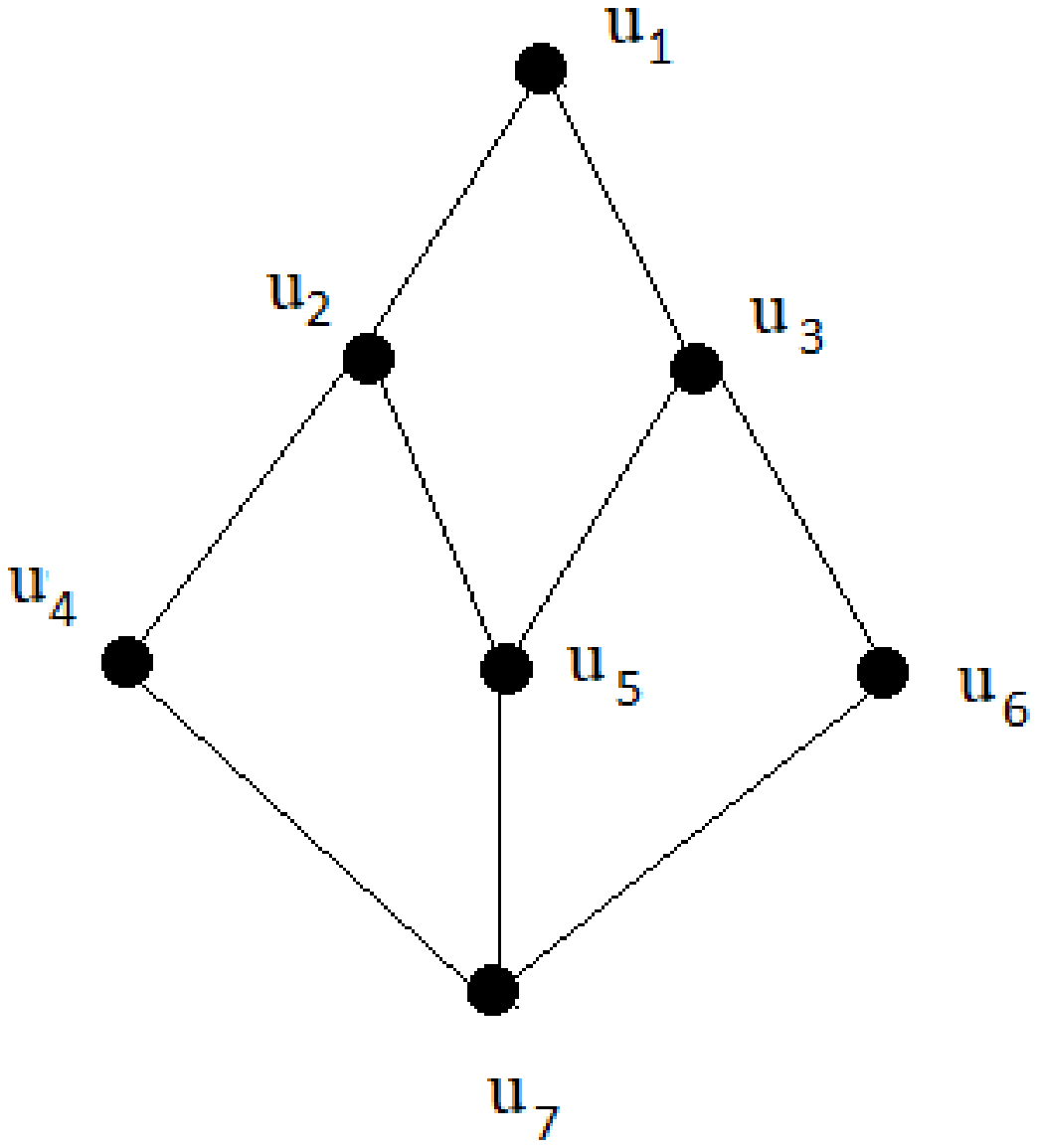}
\caption{Hierarchy of users.}
\label{fig3}
\end{figure}

By the same principle elements of the set S are calculated (the Figure 4.)

\begin{figure}[ht]
\centering
\includegraphics[width=0.5\textwidth]{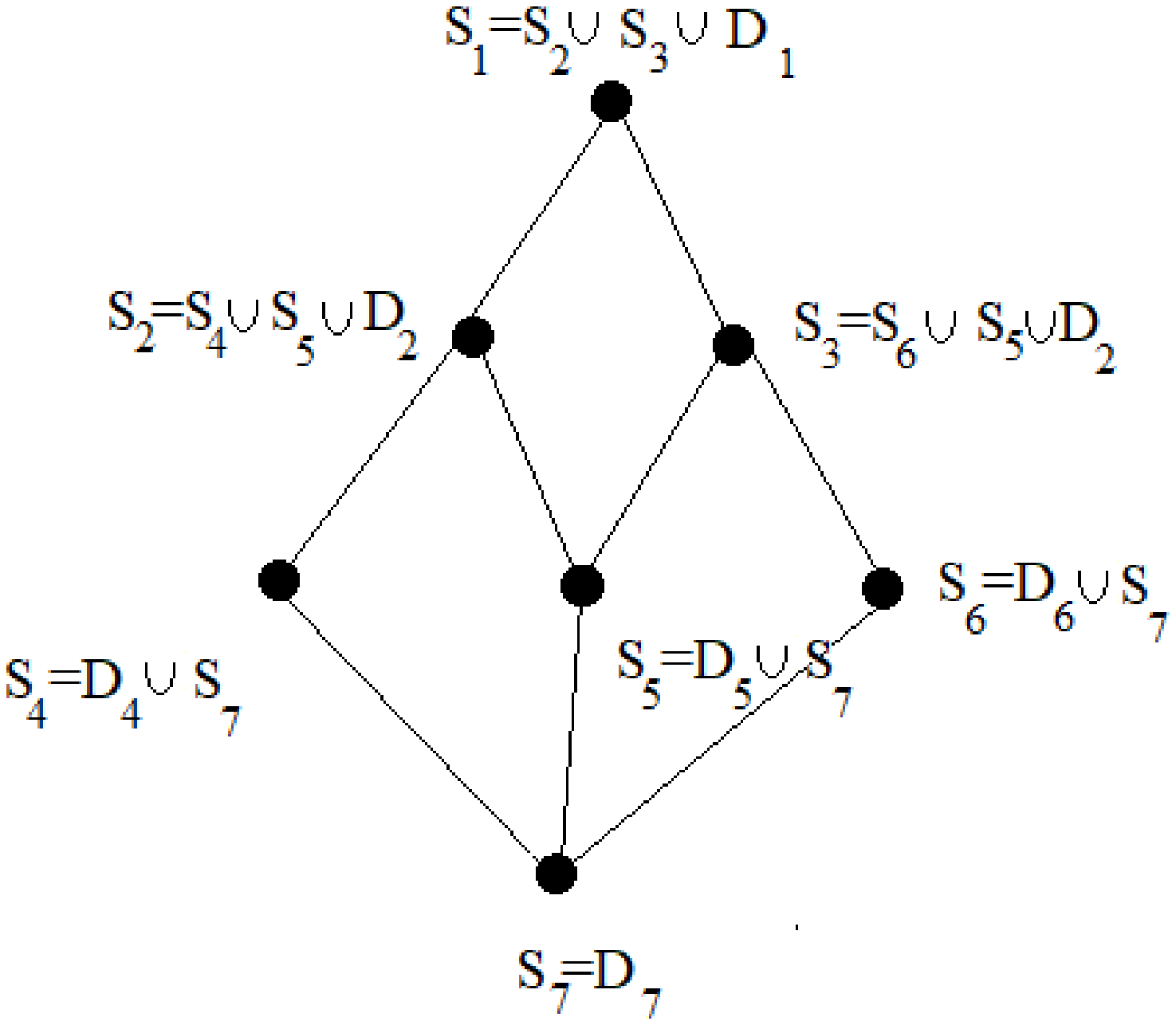}
\caption{Hierarchy of users.}
\label{fig4}
\end{figure}

The keys contradicting mandatory security policy are equal to zero.

\section*{Conclusion}

The scheme suggested in this article allows realizing keys preliminary distribution 
of a symmetric enciphering for the distributed systems with user's hierarchy. As well as 
in case of the KDP scheme the Sperner's families are used. However KDP scheme allows forming
bidirectional channels of information exchange whereas the suggested scheme is focused 
on simplex channels. The suggested modification of the keys preliminary distribution scheme 
does not increase the size of key materials. It is an indispensable condition for using this
approach to creation the protected systems.


\begin{thebibliography}{00}

\bibitem{b1}
Blom R. An optimal class of symmetric key generation systems. Proc. of the EUROCRYPT 84  pp 335-338. (1985)
\bibitem{b2}
Mitchell C.J. and Piper C  Key storage in Secure Networks Discrete and Applied Math. 21 pp 215-228. (1988)
\bibitem{b3}
Belim S.V., Belim S.Yu. and Polyakov S.Yu. The Implementation of Discretionary Access Separation
Using a Modified Blom's Scheme of Key Distribution. Information Security Problems. Computer 
Systems, 3, pp. 72--76. (2015)
\bibitem{b4}
Belim S.V. and Belim S.Yu. The Modification of Blom's Key Predistribution Scheme, Taking into
Account Simplex Channels. Information Security Problems. Computer Systems, 3, pp. 82--86. (2017)
\bibitem{b5}
Belim S.V. and Belim S.Yu. KDP Scheme of Preliminary Key Distribution in Discretionary Security
Policy.  Automatic Control and Computer Sciences, 50, 8, pp.777--786. (2016)
\bibitem{b6}
Belim S.V. and Belim S.Yu. The VPN Implementation on Base of the KDP-Scheme. CEUR Workshop
Proceedings 1732. URL: http://ceur-ws.org/Vol-1732/paper3.pdf. (2016)
\bibitem{b7}
Belim S.V. and Bogachenko N.F. Distribution of Cryptographic Keys in Systems with a Hierarchy of
Objects.  Automatic Control and Computer Sciences, 50, 8,  pp.773--776. (2016)
\bibitem{b8}
Dyer M., Fenner T., Frieze A. and Thomason A. On key storage in secure networks. 
J. Cryptology, 8, pp. 189--200. (1995)
\bibitem{b9}
O'Keefe C.M. Applications за finite geometries in information security. Australas. J. Combin.  
7, pp. 195--212. (1993)
\bibitem{b10}
Young M. The Technical Writer's Handbook. Mill Valley, CA: University Science, 1989.

\end{thebibliography}
\end{document}